\documentclass[preprint,12pt]{elsarticle}

\usepackage{graphicx}
\usepackage{pgfplots}
\usepackage[percent]{overpic}
\usepackage{caption}
\usepackage{soul}
\usepackage{subcaption}

\usepackage{amssymb}
\usepackage{amsmath}

\usepackage[a4paper, total={6.5in, 9in}]{geometry}
\usepackage{float}

\journal{Composites Part B}

\begin{document}

\begin{frontmatter}


\title{Instabilities driven by frontal polymerization in thermosetting polymers and composites}



\author[CEE,Beckman]{E. Goli}
\author[AE,Beckman]{S. R. Peterson}
\author[AE,Beckman]{P. H. Geubelle\fnref{geubelle}}

\fntext[geubelle]{Corresponding author. Email: geubelle@illinois.edu. Tel: +1(217)244-7648}

\address[CEE]{Department of Civil and Environmental Engineering, University of Illinois, Urbana, IL 61801, USA}

\address[AE]{Department of Aerospace Engineering, University of Illinois, Urbana, IL 61801, USA}



\address[Beckman]{Beckman Institute for Advanced Science and Technology, University of Illinois, Urbana, IL 61801, USA}

\begin{abstract}
Frontal Polymerization (FP) was recently demonstrated as a faster, more energy-efficient way to manufacture fiber-reinforced thermosetting-polymer-matrix composites. FP uses heat from the exothermic reaction of the solution of monomer and initiator to generate a self-propagating polymerization front. In most cases, the polymerization front propagates in a steady fashion. However, under some conditions, the front experiences instabilities, which do affect the quality of the manufactured composite part. In this work, we use a coupled thermo-chemical model and an adaptive nonlinear finite element solver to simulate FP-driven instabilities in dicyclopentadiene (DCPD) and in carbon-fiber DCPD-matrix composites. With the aid of 1-D transient simulations, we investigate how the initial temperature and the carbon fiber volume fraction affect the amplitude and wavelength of the thermal instabilities. We also extract the range of processing conditions for which the instabilities are predicted to appear. The last part of this work investigates the effect of convective heat loss on the FP-driven instabilities in both neat resin and composite cases. 
      
\newgeometry{bottom=1cm}
\end{abstract}


\end{frontmatter}


\section{Introduction}
Traditional manufacturing approaches for the fabrication of fiber-reinforced thermosetting-polymer-matrix composites rely on the bulk polymerization of the matrix phase and often require costly, time-consuming processes where the manufactured part is subjected to long and complex pressure and temperature cycles \cite{daniel1994engineering, xu2017situ, agius2013cure, herring2011effect}. These traditional manufacturing processes are expensive in terms of time, energy, carbon footprint, and infrastructure costs associated with autoclaves and heated molds \cite{ yoo2015simulation, liu2020sustainable, henning2019fast, dirk2012engineering, hasan2019design}. To address these challenges, frontal polymerization (FP) was recently demonstrated as a faster, energy-efficient alternative to composites manufacturing \cite{robertson2018rapid2}. FP is a reaction-diffusion process involving self-propagating chemical reactions driven by exothermic polymerization initiated by a local thermal stimulus. FP has been observed in various monomers including epoxies \cite{frulloni2005numerical, mariani2004uv, mariani2007synthesis, chen2006epoxy}, acrylates \cite{goli2019frontal2, nason2005uv, ivanov2002influence}, and dicyclopentadiene (DCPD) \cite{robertson2018rapid2, goli2018frontal, vyas2019manufacturing, vyas2020frontalsubmitted, goli2020impact}. The latter system is the focus of the present study. 

Under most conditions, the polymerization front propagates in a smooth, steady fashion. However, under certain conditions, the front experiences an unsteady propagation resulting in surface patterns on the manufactured part. Four distinct modes of FP-driven instabilities have been reported in the literature: (i) planar pulsation fronts which propagate in an oscillatory fashion, (ii) spinning fronts characterized by a reaction zone moving in a spiral fashion, (iii) non-planar aperiodical fronts that randomly migrate hot spots ahead of the fronts, and (iv) fingering fronts that propagate in a random crawling fashion \cite{masere1999period, riolfo2012experimental, inamdar2007spinning}. 

Experimental observations of front pulsations were first reported by Pojman \textit{et al.} \cite{pojman1992convective} in unstirred solutions of methacrylic acid and benzoyl peroxide. The authors conjuctured that both thermal diffusion and convective fluid motion ahead of the polymerization front were responsible for the presence of pulsations. In their study, they emphasized the role played by fluid convection motivated by the large thermal expansion associated with the exothermicity of the reaction and isothermal contraction. Building on this work, other authors utilized linear stability analysis to reproduce the experimentally observed instabilities by imposing small perturbations to the hydro-thermo-chemical equations \cite{volpert1996hydrodynamic, garbey1996linear, bowden1997effect, allali2014convection}. The mathematical models used in these studies consist of a reaction-diffusion system combined with the Navier-Stokes equations to capture the fluid-convection effect. The complex system of coupled nonlinear equations is then simplified by adopting a zero-order cure kinetic model and by assuming small perturbations around the closed-form expression of the steady-state front velocity. 

Other experimental studies suggested the imbalance between characteristics times associated with thermal diffusion and chemical reaction to be responsible for the appearance of these instabilities, and considered the convective fluid effects as secondary \cite{pojman1995spin, huh2002reaction, huh2003bistability}. This line of work includes later analytical studies of the contribution of first-order reaction kinetics \cite{solovyov1997numerical}, convective heat losses \cite{spade2001linear}, and reactor geometry \cite{comissiong2004bifurcation}.

The present work also focuses on the modeling of thermo-chemical instabilities, with emphasis on capturing the effect of the reinforcing fiber content in DCPD-based composites. The analysis presented hereafter relies on a finite element solution of the transient, nonlinear, coupled, thermo-chemical model  of FP to investigate planar pulsating fronts in neat DCPD resin and carbon/DCPD composites. Emphasis is placed hereafter on capturing the effect of the initial temperature of the monomer solution and of the fiber volume fraction on the wavelength and amplitude of the thermal oscillations present in the immediate vicinity of the propagating front. Motivated by experimental observations available in the literature, we also investigate the effect of convective heat loss on the instability patterns in both neat resin and carbon/DCPD composites. 

The manuscript is organized as follows. Section \ref{numerical_near_resin} summarizes the formulation and simulation results for FP-driven instabilities in neat DCPD resin, while Section \ref{numrical_composites} focuses on the modeling of instabilities during the FP-based manufacturing of  carbon/DCPD composites. Furthermore, we present an `instability design diagram' that summarizes the extent of process conditions where FP-driven instabilities are expected to be present. Finally, Section 4 investigates the effects of convective heat loss on the FP-driven instabilities.

\section{FP-Driven Instabilities in Neat DCPD Resin: Adiabatic Analysis} \label{numerical_near_resin}
\subsection{Thermo-Chemical Model }
As shown in \cite{goli2018frontal}, the initiation and propagation of a polymerization front in an adiabatic DCPD specimen of length \textit{L} can be described by the following coupled nonlinear partial differential equations (PDEs) relating the temperature $T(x,t)$ and the degree of cure $\alpha(x,t)$:

\begin{equation}
\label{PDES}
\begin{cases}
    \kappa \dfrac{\partial^2 T}{\partial x^2}+\rho H_r \dfrac{\partial \alpha}{\partial t}=\rho C_p \dfrac{\partial T}{\partial t},\vspace{2mm}\\ 
    \dfrac{\partial \alpha}{\partial t}=A exp({-\dfrac{E}{RT}})g(\alpha),\\
g(\alpha)=(1-\alpha)^{n}\alpha^{m}\dfrac{1}{1+exp\left [ C(\alpha-\alpha_c) \right ]}.
\end{cases}
\end{equation}
In (\ref{PDES}), $\kappa$, $\rho$, $H_r$, and $C_p$ represent the thermal conductivity ($0.152~W/m.K$), density ($980~kg/m^3$), total enthalpy of reaction ($350~J/g$), and specific heat ($1600~J/kg. K$) of the DCPD resin, respectively. The second and third equations describe the cure kinetics model with $A$ ($8.55 \times 10^{15}  s^{-1}$) and $E$ ($110.75~kJ/mol$) denoting the pre-exponential factor and the activation energy, respectively, and $R$ ($8.314~J/mol. K$), the universal gas constant. The Prout-Tompkins model, $(1-\alpha)^n\alpha^m$, with exponents $m= 0.77$ and $n=1.7215$ is completed by the term $(1+exp[C(\alpha - \alpha_c)])^{-1}$, with $C=14.478$ and $\alpha_c= 0.405$, to capture the effects of diffusion at higher temperatures \cite{yang2014curing}. The six constants in the cure kinetics equation, $A,E,n,m,C$, and $\alpha_c$, are found by nonlinear fitting the rate of cure data extracted from dynamic Differential Scanning Calorimetry (DSC) tests.

The 1-D model of FP is completed by the following initial and boundary conditions:
\begin{equation}
\label{BCs}
\begin{cases} T(x,0)=T_0,
\\\alpha(x,0)=\alpha_0,
\\T(0,t)=T_{trig} \textrm{  }  \textrm{      for} \textrm{      } \textrm{      } 0\leq  t    \leq t_{trig}, \vspace{2mm}\\ \dfrac{\partial T}{\partial x}\mid_{(0,t)}=0  \textrm{  }  \textrm{  for}  \textrm{  }  \textrm{  } t > t_{trig},
\vspace{2mm}\\ \dfrac{\partial T}{\partial x}\mid_{(L,t)}=0,
\end{cases}
\end{equation}
where $t_{trig}$ is the time during which the triggering temperature, $T_{trig}$, is applied to the left side of the domain to initiate polymerization, and $\alpha_0$ denotes the initial degree of cure. Solving the thermo-chemical PDEs is complicated by the presence of sharp gradients in the temperature and cure profiles associated with the advancing front. To address this challenge, the Multiphysics Object Oriented Simulation Environment (MOOSE) is adopted in this work. MOOSE is an open source, C++ math library that provides a robust h-adaptivity module which monitors the gradients associated with the degree of cure and temperature solutions and adjusts the finite element mesh accordingly  \cite{permann2020moose}. Implicit Euler time integration and Preconditioned Jacobian-Free Newton-Krylov  schemes are used to solve the nonlinear system of equations at each time step \cite{goli2018frontal, goli2019frontal2}, and an adaptive time stepping algorithm is implemented to reduce the cost of computation.

\begin{figure}[H]
    \centering
    \includegraphics[scale=0.7]{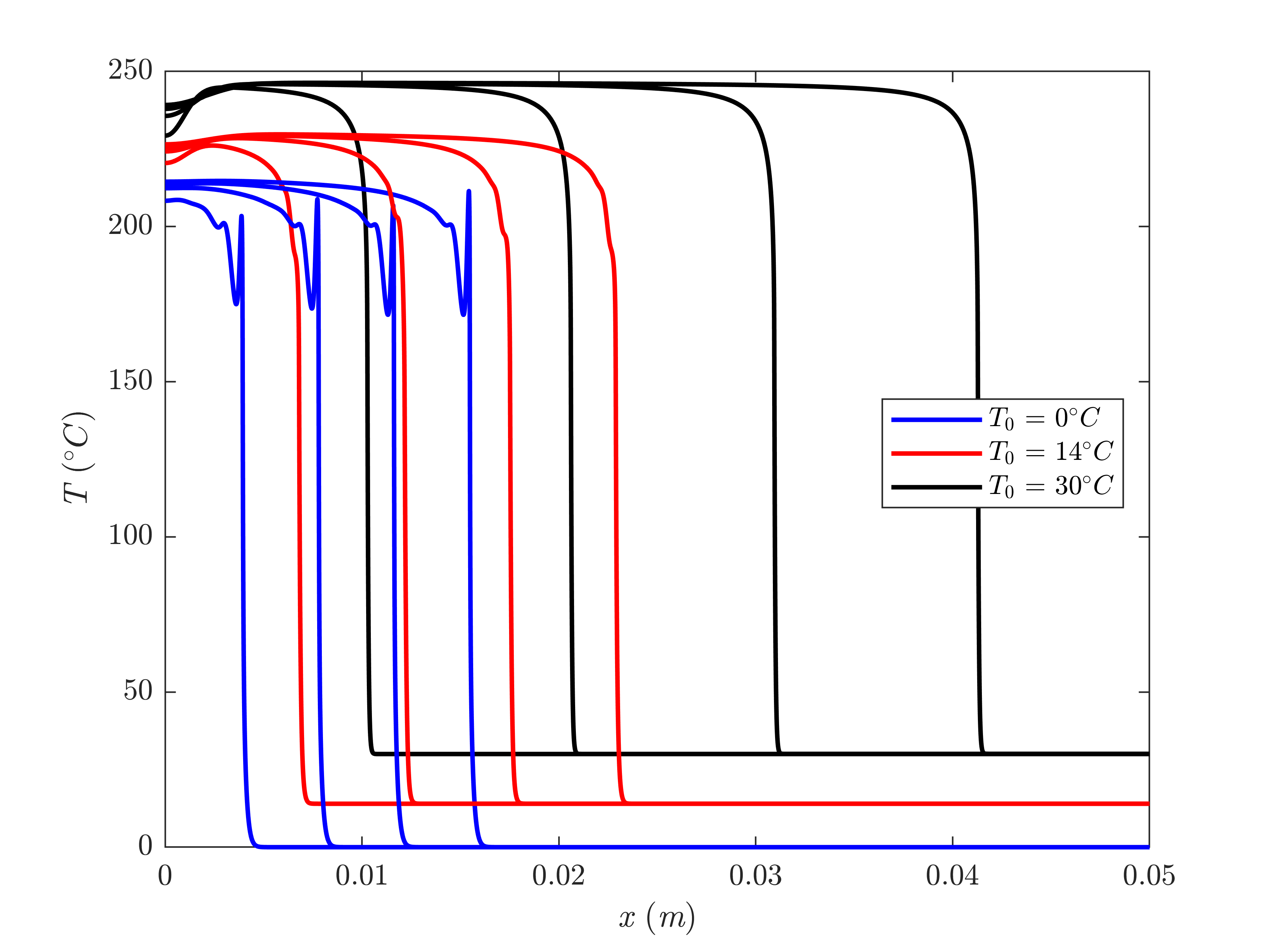}
    \caption{Temperature profile at different initial temperatures at $t = 5, 10, 15,$ and $20 s$. Higher initial temperatures improve the stability of the fronts. }
    \label{fig:5}
\end{figure}

\subsection{Numerical Results} \label{results_neat_resin}
 To recreate instabilities numerically, a 1-D reaction channel of length $L=5cm$ filled with a DCPD monomer solution is modeled at different initial temperatures ranging from $T_0 = 0^{\circ}C$ to $T_0 = 30 ^{\circ}C$ using a uniform mesh consisting of 1250 nodes before mesh adaptivity module is activated. The initial degree of cure, $\alpha_0$, is set at 0.01 and we initiate the front with $t_{trig}=1s$ and $T_{trig}=200^oC$. Figure \ref{fig:5} depicts the thermal solution corresponding to three values of the initial temperature at $t = 5, 10, 15,$ and $20 s$. 
 
 \begin{figure}[H]
    \centering
    \includegraphics[scale=0.7]{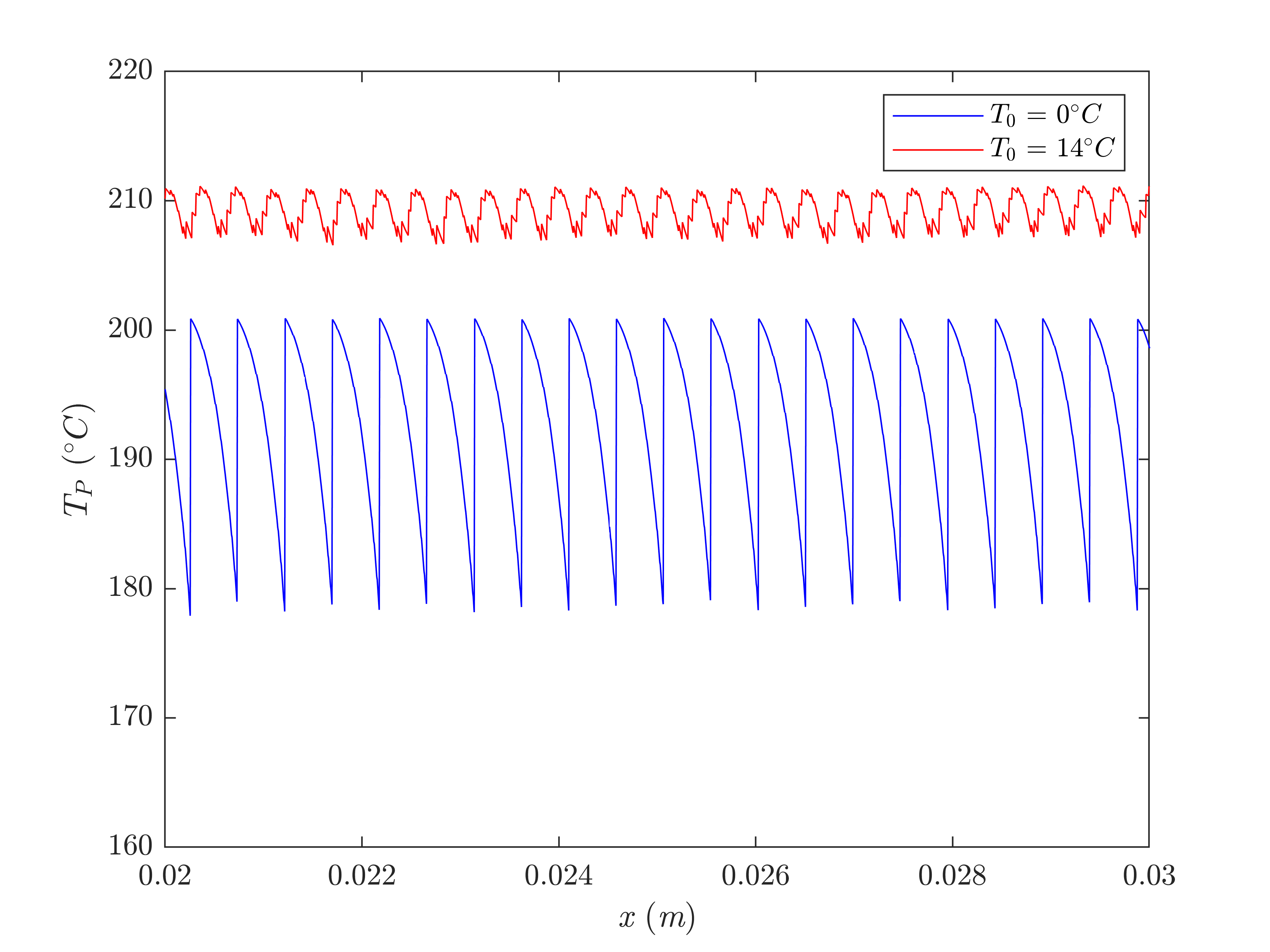}
    \caption{Spatial variation of the thermal spike obtained for $T_0 = 0^{\circ}C$ and $T_0 = 14^{\circ}C$, showing a sharp dependence of the amplitude and wavelength of thermal spikes on the initial temperature of the monomer.}
    \label{fig:6}
\end{figure}

At lower initial temperatures, the front experiences a repeatable sharp thermal spike prior to achieving its final temperature as it propagates through the reaction channel. However, as the initial temperature increases, the amplitude of these thermal spikes progressively reduces until a smoothly propagating front is obtained as illustrated for the case $T_0 = 30^{\circ}C$, as shown in Figure \ref{fig:5}. The figure also illustrates how the front speed increases with increasing values of $T_0$.

\begin{figure}
\begin{subfigure}{.5\textwidth}
\centering
\includegraphics[width=1\linewidth]{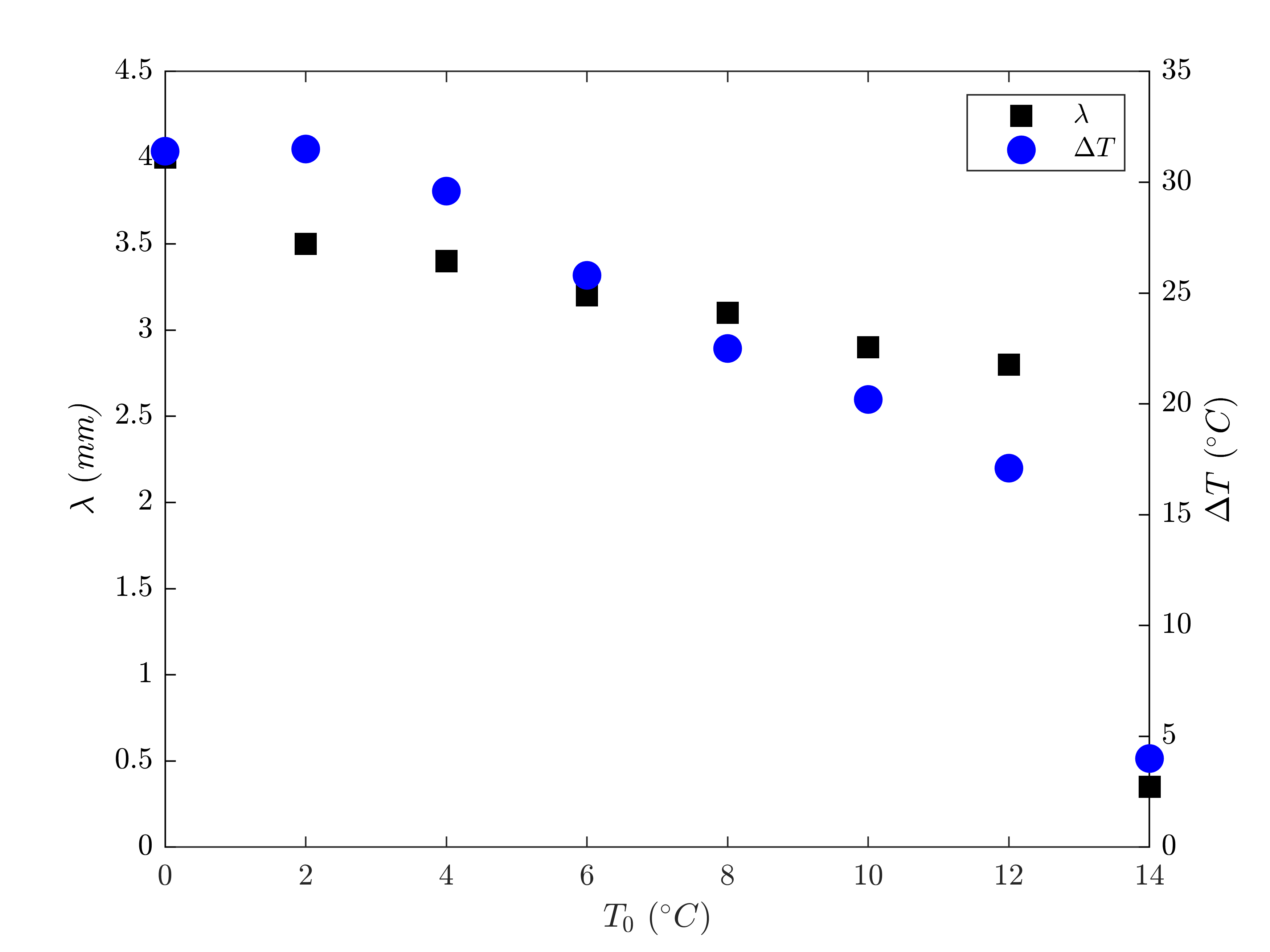}
\subcaption{}
\end{subfigure}
\begin{subfigure}{.5\textwidth}
\centering
\includegraphics[width=1\linewidth]{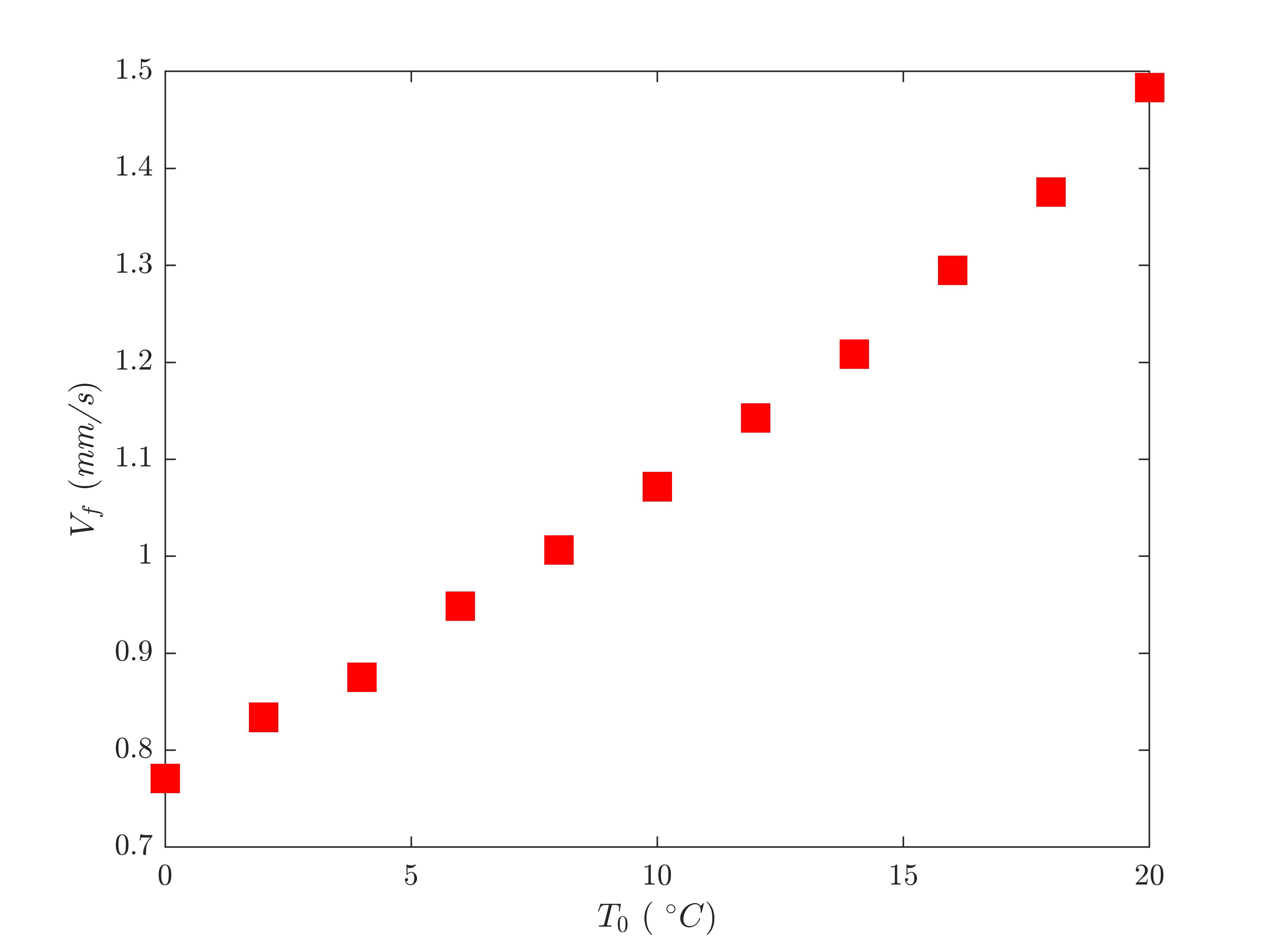}
\subcaption{}
\end{subfigure}
\caption{(a) Dependence of the wavelength and amplitude of thermal instabilities on the initial temperature $T_0$. (b)  Front velocity in neat DCPD resin as a function of $T_0$ showing that the front velocity is not affected by the thermal instabilities taking place for $T_0 \leq 14^{\circ}C$.}
	\label{neat_resin_mix}
\end{figure}

For the system of interest, the thermal instabilities are found to arise for $T_0 \leq 14^{\circ}C$. Extracted from bifurcation analysis, the non-dimensional Zeldovich number has been frequently used in the literature \cite{volpert1996hydrodynamic, garbey1996linear, bowden1997effect, solovyov1997numerical} as the key measure for the appearance of pulsating instabilities:
\begin{equation}
\label{Zeldovic}
Z=\frac{T_{max}-T_0}{T_{max}}\dfrac{E}{RT_{max}}.
\end{equation}
 Given an adiabatic system where the maximum temperature can be analytically obtained by $T_{max}= T_0+H_r(1-\alpha_0)/C_p$, the value of the Zeldovich number corresponding to the onset of instabilities is found to be 11.37. In other words, for $Z\leq 11.37$, the thermo-chemical instabilities are not expected to appear in DCPD resin frontally polymerized in a perfectly insulated 1-D reactor.

To characterize the thermal instabilities illustrated in Figure \ref{fig:5}, the values of the maximum temperature ($T_P$) associated with the thermal spike at the passage of the front are presented in Figure \ref{fig:6}, showing the repeatability of the instability process and the strong dependence of the amplitude and wavelength of the thermal spikes on the initial temperature of the monomer.

Figure \ref{neat_resin_mix}(a) depicts the dependence of the wavelength ($\lambda$) and amplitude ($\Delta T$) of the computed pulsating instability on the initial temperature ($T_0$) of the DCPD monomer.
As apparent in that figure, both $\Delta T$ and $\lambda$ progressively decrease as the initial temperature increases, indicating increasingly more stable fronts, especially for $T_0 > 12^{\circ}C$ at which a sharp drop of both $\Delta T$ and $\lambda$ is observed.

It should be noted, however, that the thermal instabilities do not appear to affect the average front velocity. As shown in Figure \ref{neat_resin_mix}(b), the front velocity consistently increases with $T_0$ and appears to be unaffected by the presence of the thermal instabilities occuring for $T_0 \leq 14^{\circ}C$.

\section{FP-Driven Instabilities in Carbon/DCPD Composite: Adiabatic Analysis} \label{numrical_composites}
\subsection{Homogenized Thermo-Chemical Model}
Instabilities also arise in the composite system and produce undesirable surface patterns. In recent studies \cite{vyas2019manufacturing, goli2019frontal3, ICCM2019}, it was proposed and validated that the following set of PDEs capture the evolution of temperature and degree of cure during FP-based manufacturing of fiber-reinforced polymer composites:

\begin{equation}
\label{comp_inst}
\begin{cases}
    \bar\kappa \dfrac{\partial^2 T}{\partial x^2}+\bar\rho H_r(1-\phi) \dfrac{\partial \alpha}{\partial t}=\bar\rho \bar C_p \dfrac{\partial T}{\partial t},\\
    \dfrac{\partial \alpha}{\partial t}=A exp({-\dfrac{E}{RT}})(1-\alpha)^{n}\alpha^{m}\dfrac{1}{1+exp\left [ C(\alpha-\alpha_c) \right ]},
\end{cases}
\end{equation}
where $\phi$ denotes the volume fraction of the fibers. The overbars indicate the effective (homogenized) properties of the composite, which, for the case of the unidirectional fibers in an 1-D setting, can be described by the rule of mixtures:

\begin{equation}
    \begin{cases}
    \bar\kappa = \kappa_m(1-\phi) + \kappa_f\phi, \\
    \bar\rho = \rho_m(1-\phi) + \rho_f\phi, \\
    \bar C_p = C_{pm}(1-\phi) + C_{pf}\phi. \\
    \end{cases}
\end{equation}
The subscripts $m$ and $f$ refer to the matrix and fiber properties, respectively, and the material properties of the carbon fibers are $\kappa_f = 9.3~W/m.K$, $\rho_f =1800~kg/m^3$, and $C_{pf}=754.6~J/kg.K$. 

\subsection{Numerical Results} \label{results_composites}
\begin{figure}[H]
    \centering
    \includegraphics[scale=0.45]{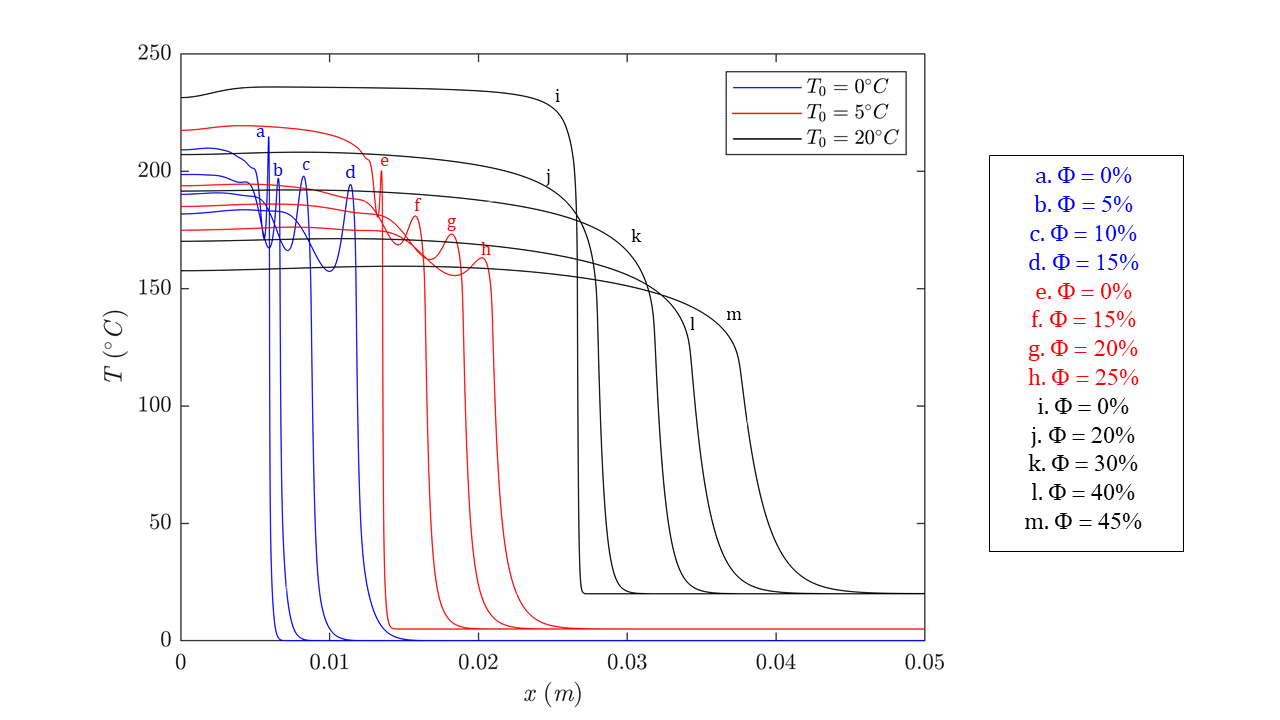}
    \caption{Temperature profile at different values of $\phi$ and $T_0$, showing that composites with lower fiber volume fractions are more prone to instabilities.}
    \label{fig:8}
\end{figure}

To simulate front instabilities in DCPD/carbon fiber composites, a 1-D reaction channel of $5~cm$ with 1250 nodes filled with the DCPD resin system and unidirectional carbon fibers is modeled with fiber volume fractions ranging from $\phi = 0\%$ to $\phi = 50\%$, and with $0\leq T_{0}\leq 30 ^\circ C$. The boundary conditions, including the thermal triggering of the FP, are the same as those used in the previous section. Figure \ref{fig:8} depicts the solution to the system of PDEs for varying values of $\phi$ and $T_0$.

\begin{figure}[H]
    \centering
    \includegraphics[scale=0.6]{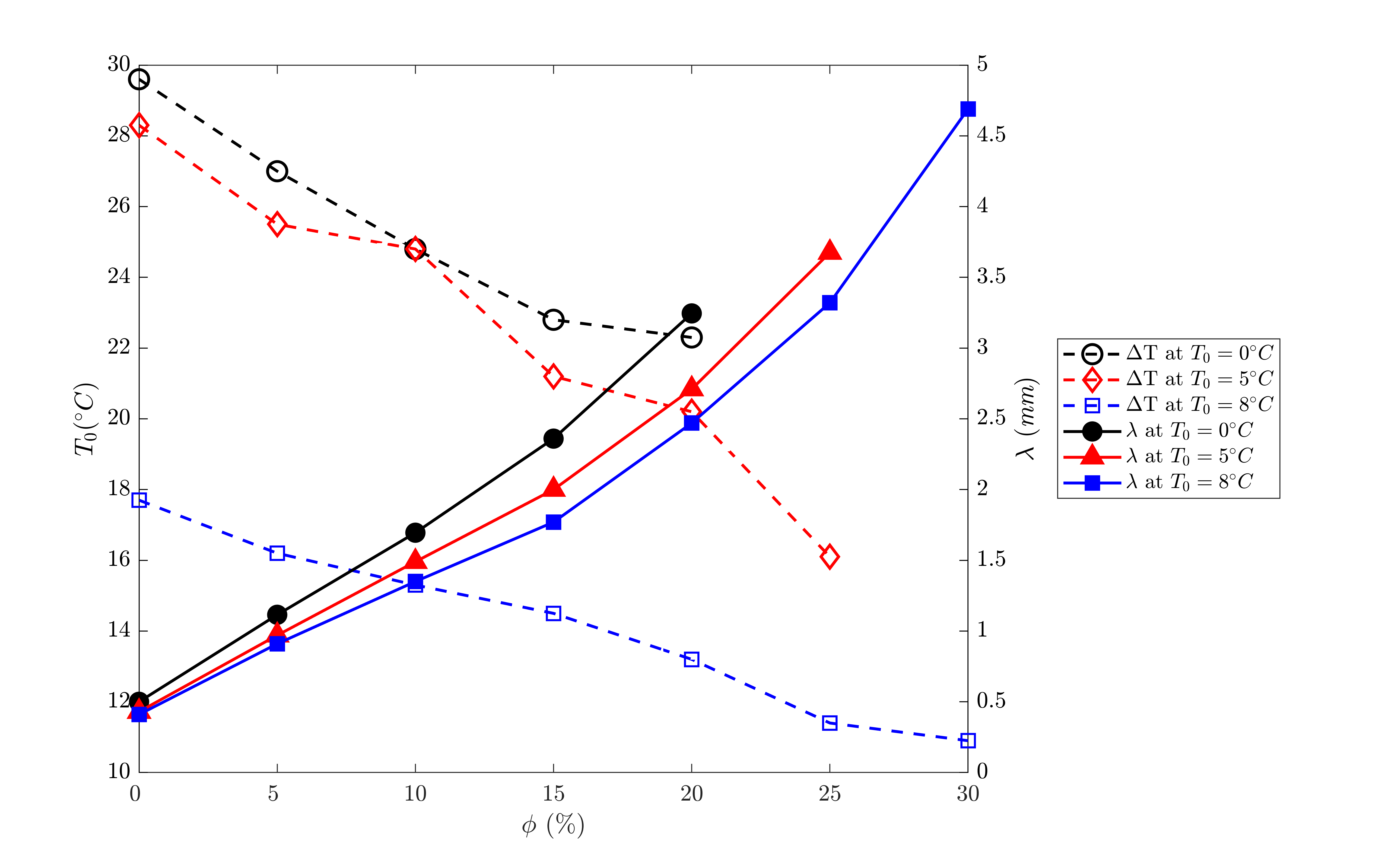}
    \caption{Amplitude, $\Delta T$, and wavelength, $\lambda$, of instabilities as functions of the fiber volume fraction, $\phi$, for three values of the initial temperature, $T_0$.}
    \label{fig:9}
\end{figure}

As with the neat resin system, the amplitude ($\Delta T$) and wavelength ($\lambda$) of the instabilities are extracted from the temperature profiles computed for the carbon/DCPD composite (Figure \ref{fig:9}).
As shown in Figure \ref{fig:9}, $\Delta T$ decreases with the fiber volume fraction, while the wavelength $\lambda$ increases with the fiber volume fraction due to the presence of conductive carbon fibers in the composite system. The increase in $\lambda$ coupled with the decrease in $\Delta T$ leads to a more stable system.

\subsection{Instability Design Space} \label{design_space}
\begin{figure}[H]
    \centering
    \includegraphics[scale=0.8]{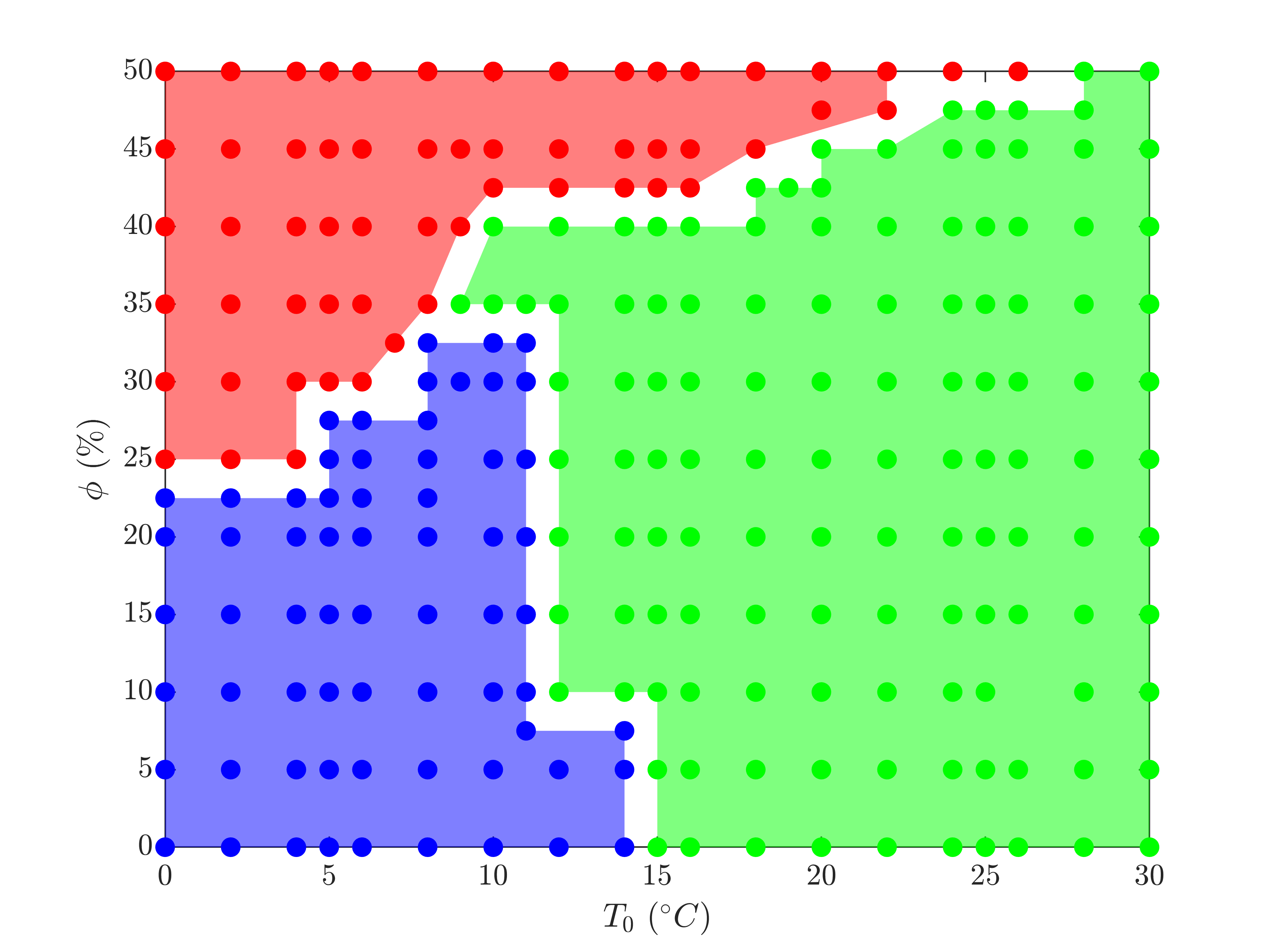}
    \caption{Manufacturing design diagram, showing how the combined effect of initial temperature and fiber volume fraction leads to stable, unstable, or quenching polymerization fronts.}
    \label{fig:10}
\end{figure}

To summarize the parametric neat resin and composite studies in adiabatic condition, we present in Figure \ref{fig:10} a `manufacturing design diagram' by classifying the predicted propagation response of the polymerization front in the ($\phi$, $T_0$) space as \textit{quenched}, \textit{unstable}, or \textit{stable}. 
`Quenched' refers to a polymerization front that fails to propagate through the entire reaction channel, `unstable' refers to combinations of $\phi$ and $T_0$ for which instabilities arise, and `stable' refers to a smoothly propagating front.
The upper-left red region denotes quenched fronts associated with the amount of heat input described in Section \ref{results_composites}. In these cases, the diffusion mechanism dominates and the temperature profile created by the thermal triggering drops immediately after removing $T_{trig}$ and the front stops. The lower-left blue region represents conditions in which pulsating instabilities arise, i.e., for which the front propagates in an unstable manner due to the interplay in the characteristic times associated with diffusion and reaction mechanisms. The right-most green section denotes combinations of $\phi$ and $T_0$ that yield a smoothly propagating front.

\section {FP-Driven Instabilities in Neat Resin and Carbon/DCPD Composites: Effect of Convective Heat Loss}
The thermal equilibrium taking place between the heat generated by the exothermic curing of the resin and the thermal diffusion occurring ahead of the propagating front is essential to ensure the stability of the front. In the previous sections, we illustrated how lowering the ambient temperature adversely affects this equilibrium. In this section, we investigate the impact of a convective heat loss on this thermal equilibrium and the resulting thermal instabilities. To that effect, we modify Equation (\ref{comp_inst}) to capture the effect of heat sink by adding the convective heat loss term $\dfrac{hP}{S}\left ( T -T_0\right)$ \cite{lienhard2019heat, aragon2011computational} as:
\begin{equation}
\label{PDES_heat_sink}
\begin{cases}
    \bar\kappa \dfrac{\partial^2 T}{\partial x^2}+\bar\rho H_r(1-\phi) \dfrac{\partial \alpha}{\partial t}=\bar\rho \bar C_p \dfrac{\partial T}{\partial t}+\dfrac{hP}{S}\left ( T -T_0\right ), \vspace{1mm}\\
    \dfrac{\partial \alpha}{\partial t}=A exp({-\dfrac{E}{RT}})(1-\alpha)^{n}\alpha^{m}\dfrac{1}{1+exp\left [ C(\alpha-\alpha_c) \right ]},
\end{cases}
\end{equation}
where $h$ (in $W/m^2.K$) is the film coefficient, while $P$ (in $m$) and $S$ (in $m^2$) denote the the perimeter and the area of the channel's cross-section, respectively.

\subsection{Numerical Results for Neat Resin}
The parametric study presented hereafter is similar to that described in Section 2, with the same initial and boundary conditions except that the triggering time, $t_{trig}$ is increased to 2$s$ to ensure that enough energy is delivered to initiate the front in the presence of the heat sink. A rectangular cross section with the dimensions of 1 $cm$ $\times$ 1 $mm$ is adopted for the reaction channel and the film coefficient is set to 50 $W/m^2.K$, resulting in $hP/S$ = 110 $kW/m^3.K$.

As shown in Figure 7, which presents the evolution of the normalized temperature $\theta$ and the degree of cure $\alpha$ along the channel for different initial temperature of the resin, the addition of the convective heat loss has a significant impact on the computed thermal instabilities. As before, the normalized temperature $\theta$ is defined as

\begin{equation}
\label{thetatheta}
\theta = \frac{T-T_0}{T_{max}-T_0},
\end{equation}
where $T_{max} = T_0+H_r(1-\alpha_0)/C_p$. As apparent in Figure 7, the FP-driven thermal instabilities become more chaotic, with spatially varying amplitude and wavelength. Instabilities are observed for higher values of the resin's initial temperature than for the adiabatic case discussed in Section 2. As expected, the temperature drop behind the front due to the convective heat loss lowers the maximum temperature of the front. Note however the substantial variations in the maximum temperature $T_P$ obtained at every point, especially for the case $T_0=5C^o$ (Fig. 7(b)), for which $T_P$ ranges from 175$^oC$ to 248$^oC$. As the initial temperature increases, the maximum temperature achieved at each location follows a more chaotic distribution albeit with a reduced amplitude.

\begin{figure}
\begin{subfigure}{.5\textwidth}
\centering
\includegraphics[width=1\linewidth]{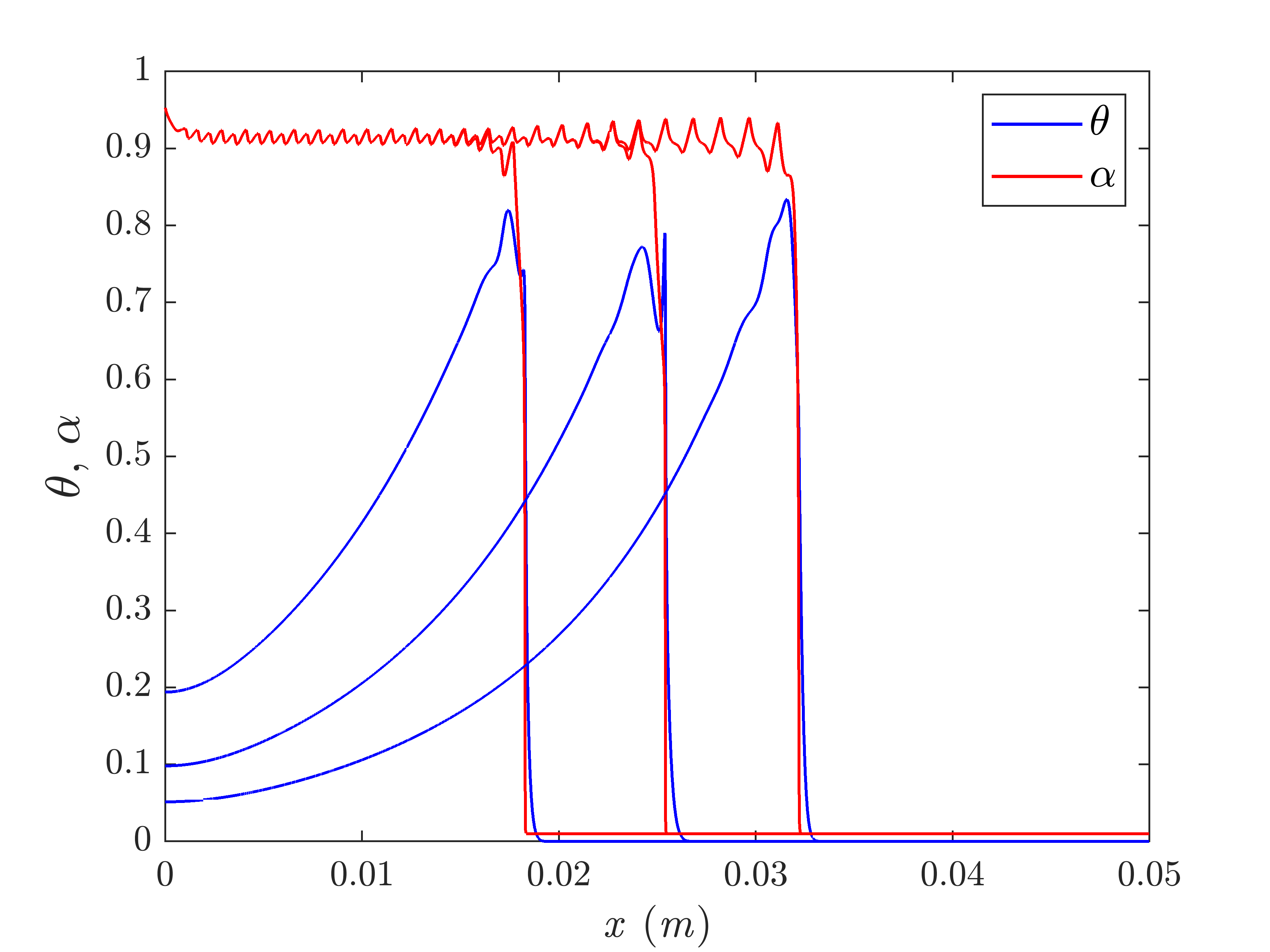}
\subcaption{}
\end{subfigure}
\begin{subfigure}{.5\textwidth}
\centering
\includegraphics[width=1\linewidth]{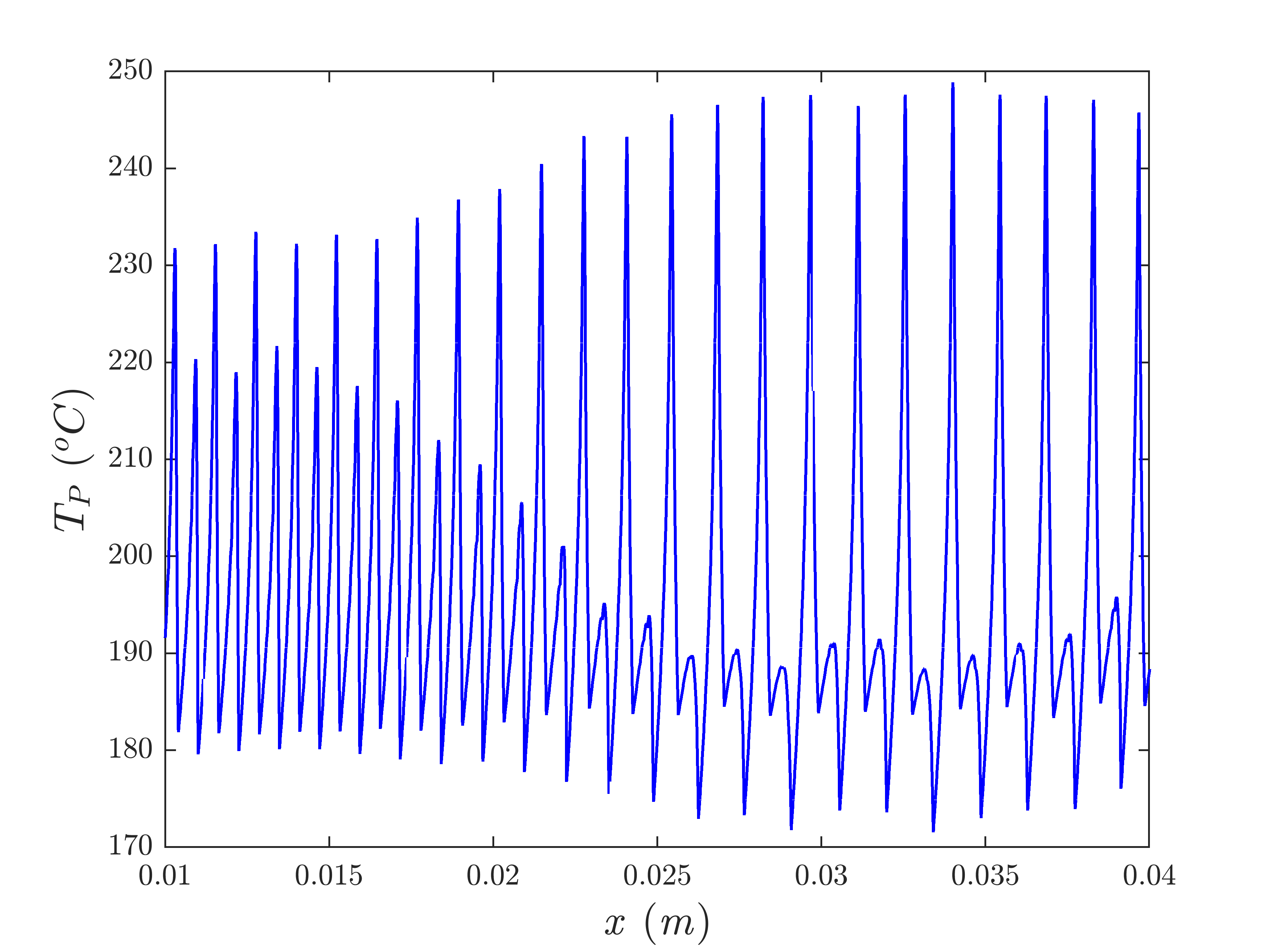}
\subcaption{}
\end{subfigure}
\begin{subfigure}{.5\textwidth}
\centering
\includegraphics[width=1\linewidth]{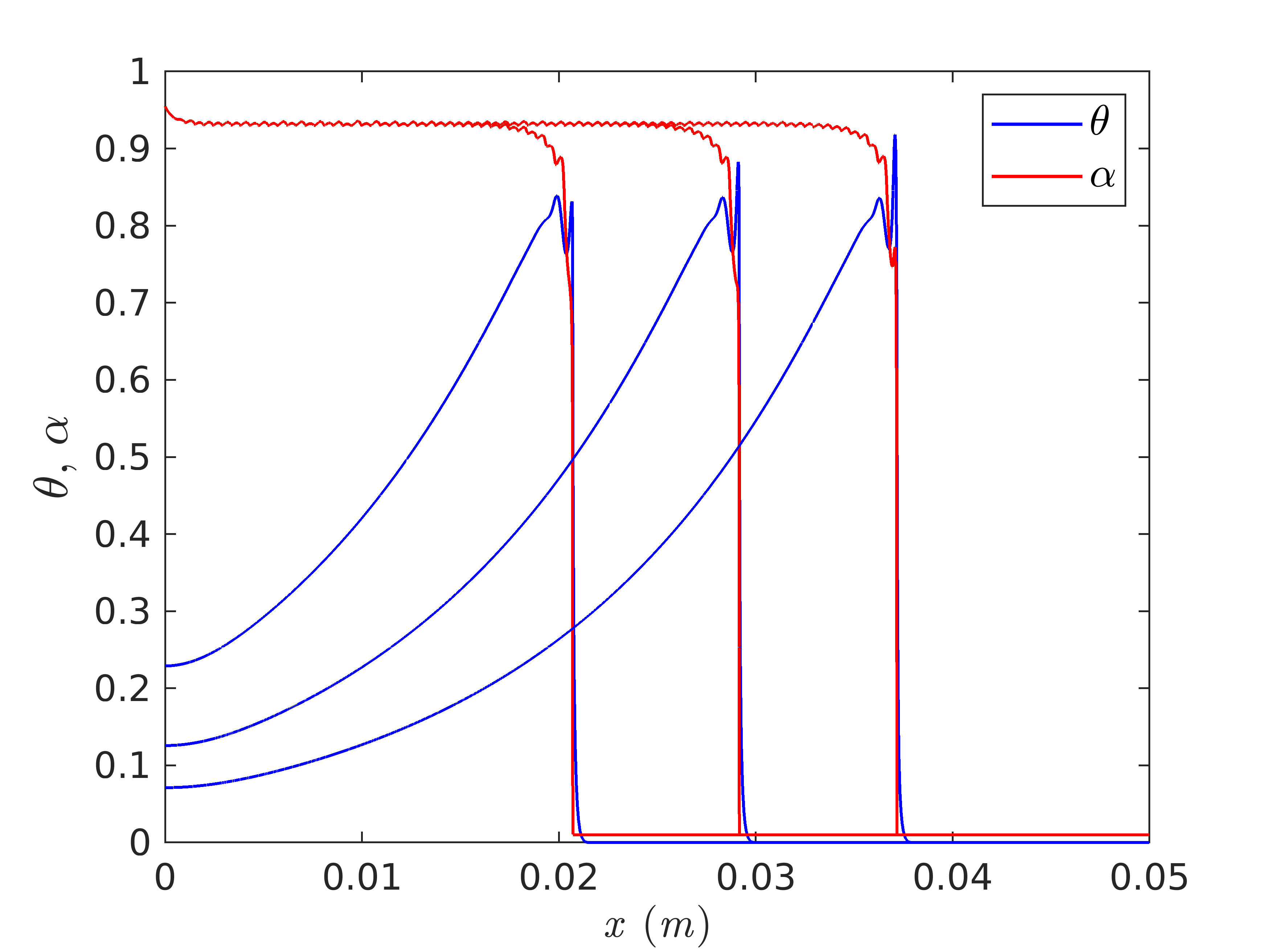}
\subcaption{}
\end{subfigure}
\begin{subfigure}{.5\textwidth}
\centering
\includegraphics[width=1\linewidth]{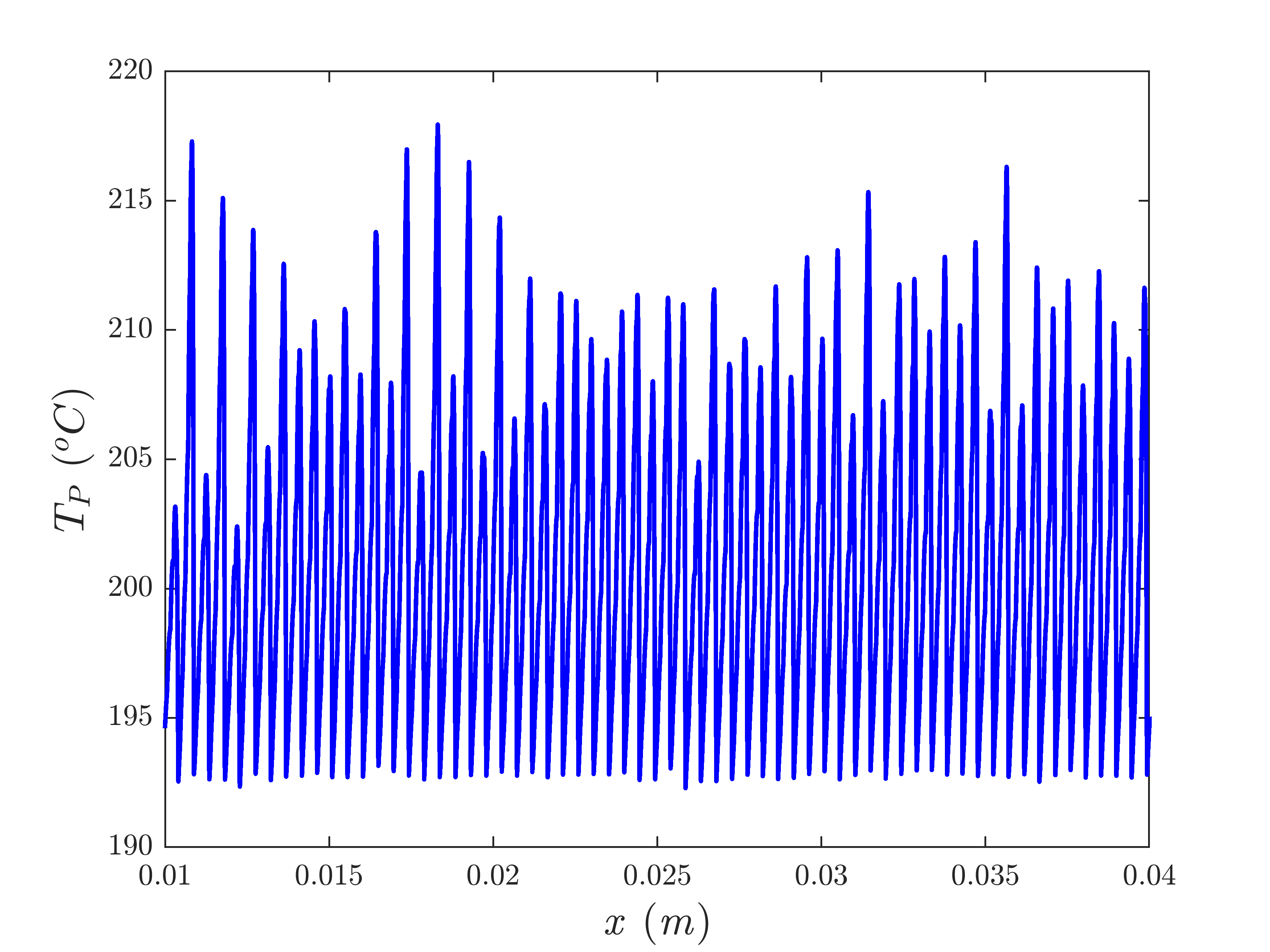}
\subcaption{}
\end{subfigure}
\begin{subfigure}{.5\textwidth}
\centering
\includegraphics[width=1\linewidth]{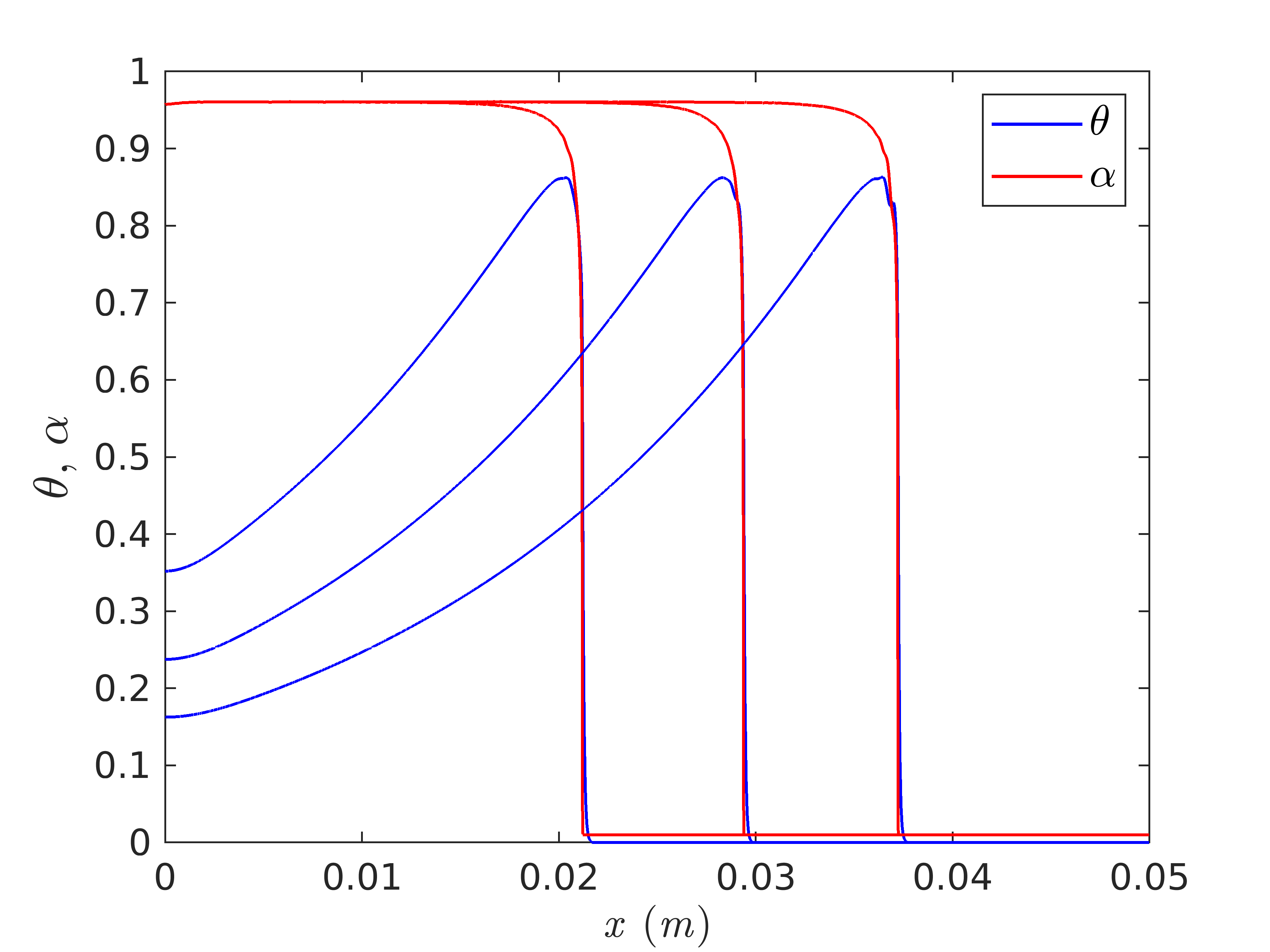}
\subcaption{}
\end{subfigure}
\begin{subfigure}{.5\textwidth}
\centering
\includegraphics[width=1\linewidth]{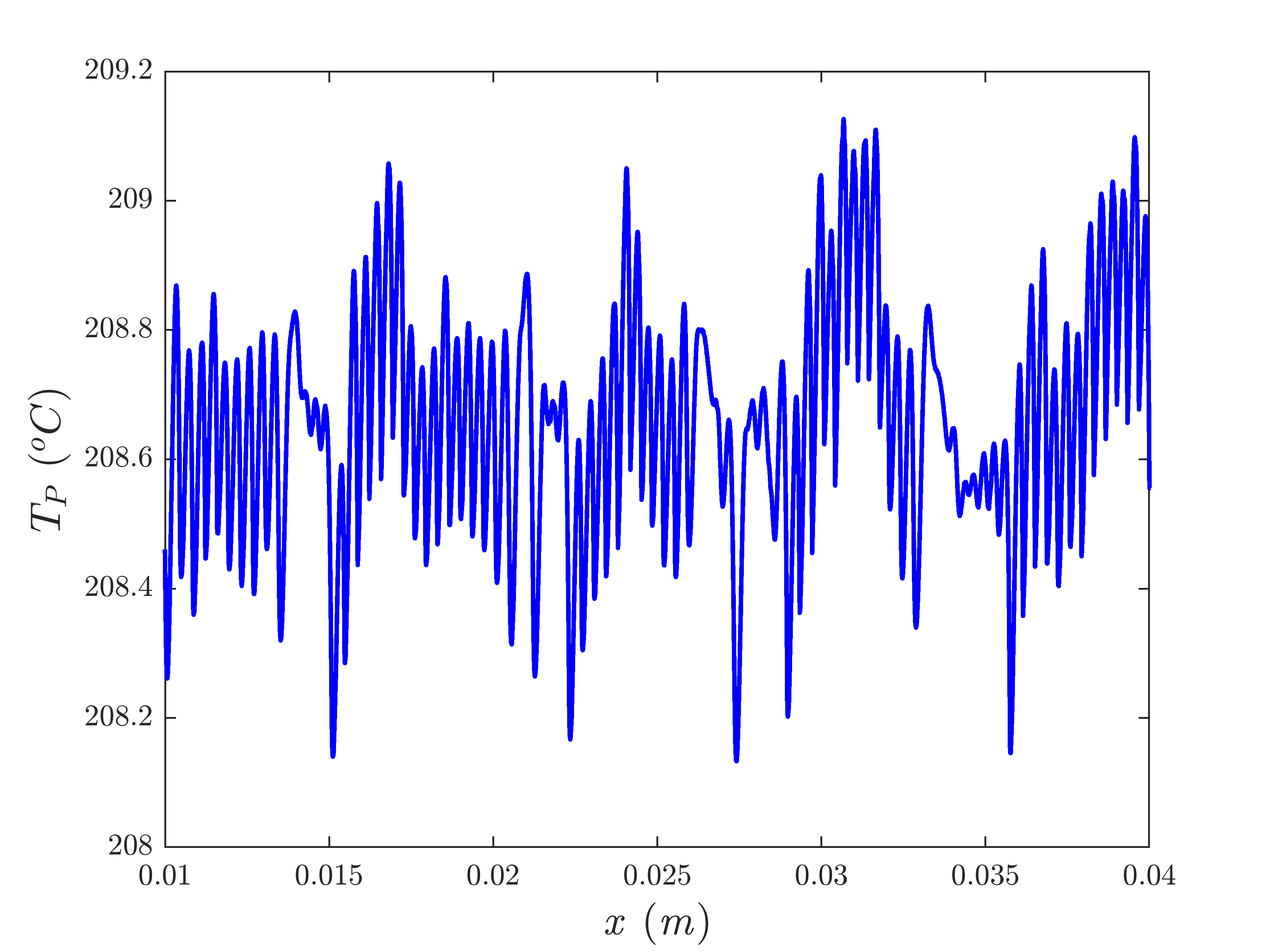}
\subcaption{}
\end{subfigure}
\caption{Normalized temperature $\theta$ and degree of cure $\alpha$ profiles for the neat resin cases with (a) $T_0=$ 5$^oC$, (c) $T_0=$ 10$^oC$, and (e) $T_0=$ 20$^oC$. showing a more complex instability pattern over a wider range of the resin's initial temperature than in the adiabatic case. Corresponding maximum temperature experienced by each point along the reaction channel during the polymerization for the neat resin cases with (b) $T_0=$ 5$^oC$, (d) $T_0=$ 10$^oC$, and (f) $T_0=$ 20$^oC$. In contrast to the harmonic instabilities presented for the adiabatic conditions (Fig. 2), spatially varying instabilities with chaotic patterns are captured.}
	\label{neat_T_alpha}
\end{figure}

\begin{figure}[H]
    \centering
    \includegraphics[scale=0.9]{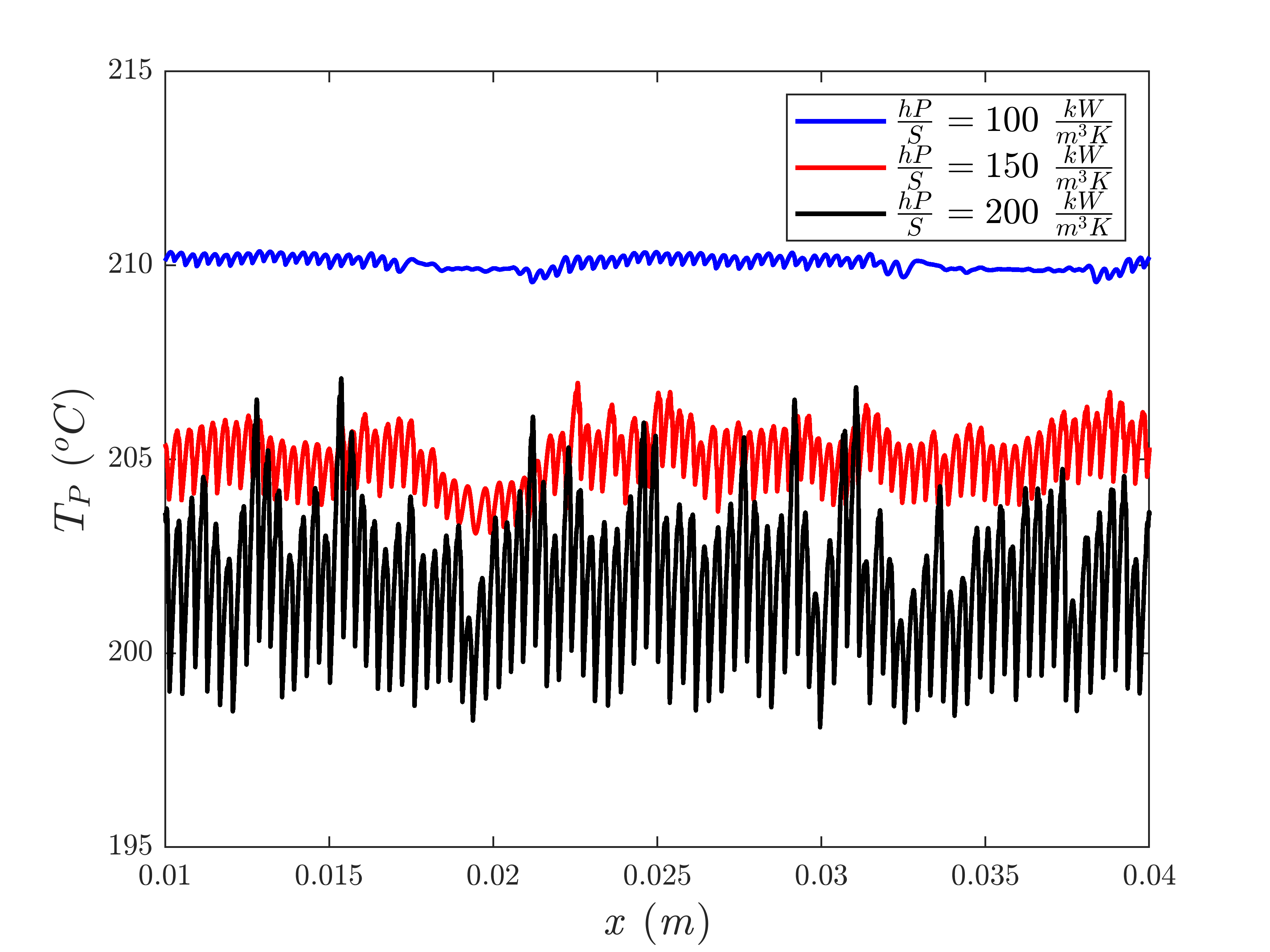}
    \caption{Maximum temperature profiles for $\dfrac{hP}{S}=$ 100, 150, and 200 $\dfrac{kW}{m^3K}$ obtained for $T_0=$ 20$^oC$. Decreasing the convective heat loss raises the peak temperatures and mitigates the intensity of the thermal instabilities.}
    \label{rates_mix}
\end{figure}

Figure \ref{rates_mix} presents the spatial variation of the peak temperature $T_P$ obtained for different channel cross-sections and/or film coefficients, i.e., for $hP/S=$100, 150, and 200 $kW/m^3.K$. As apparent there, higher convective losses lead to stronger and more complex instabilities and to lower values of the peak temperature.

\subsection{Numerical Results for Composites}
\begin{figure}
\begin{subfigure}{.5\textwidth}
\centering
\includegraphics[width=1\linewidth]{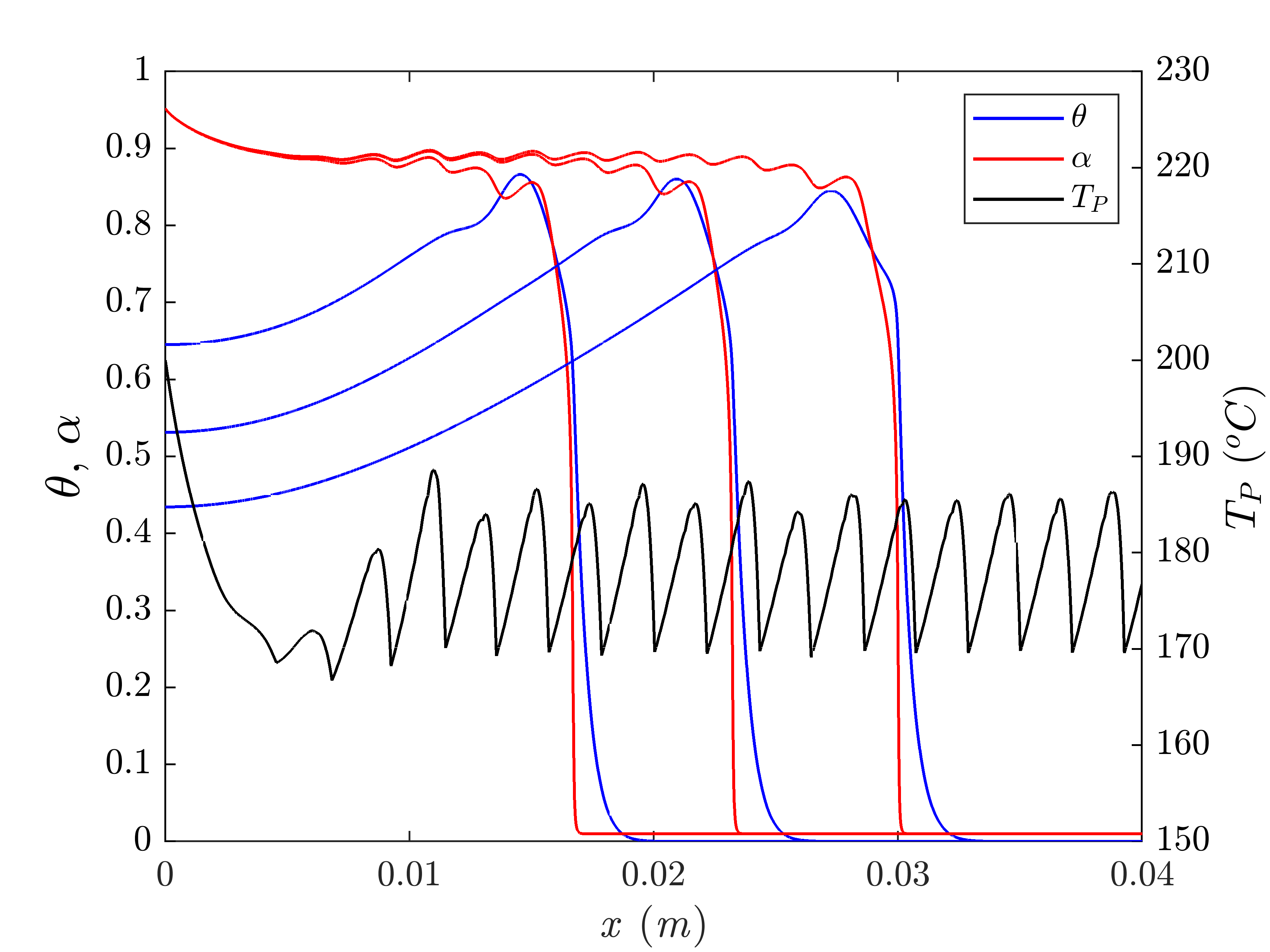}
\subcaption{}
\end{subfigure}
\begin{subfigure}{.5\textwidth}
\centering
\includegraphics[width=1\linewidth]{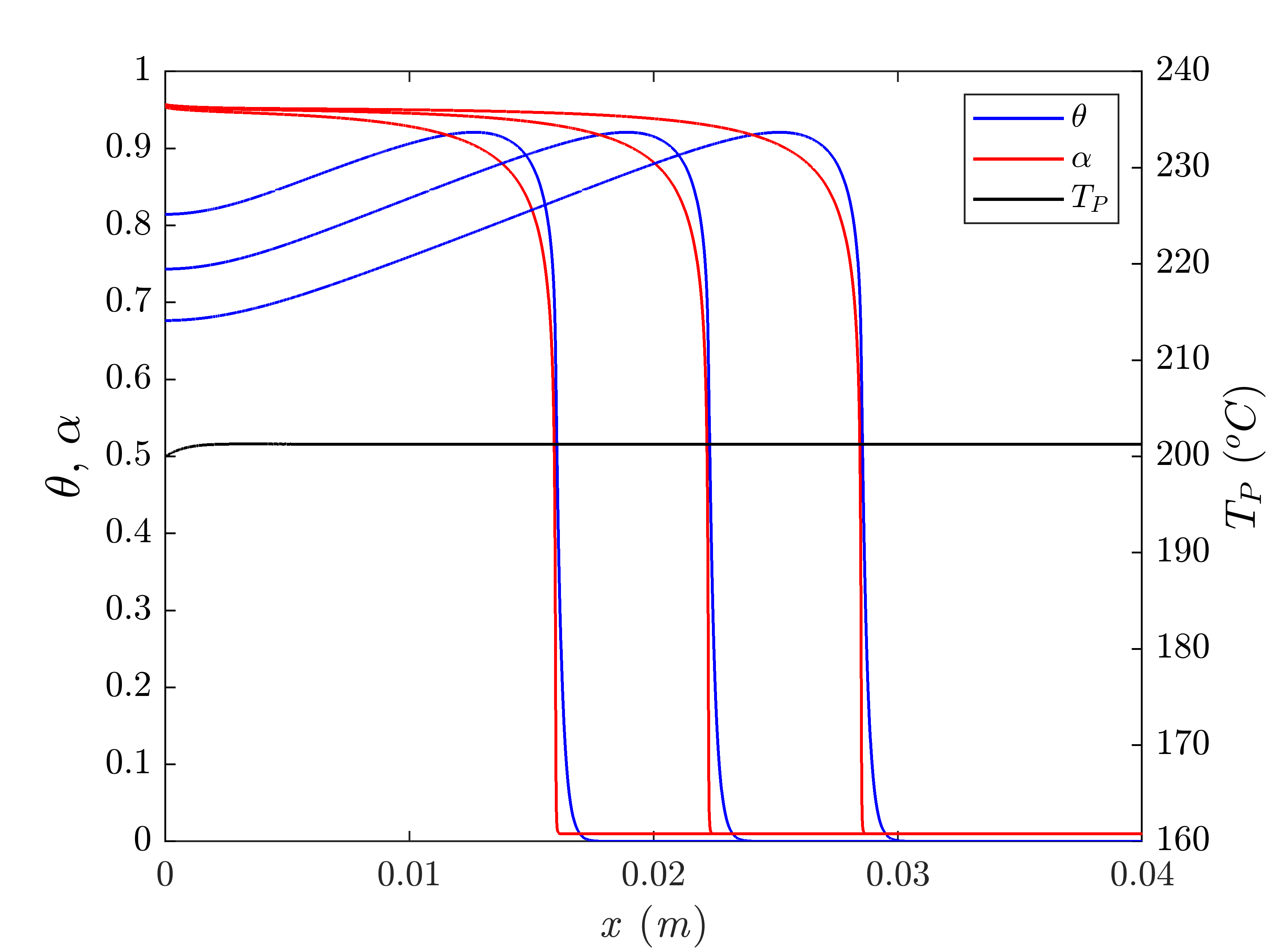}
\subcaption{}
\end{subfigure}
\caption{Normalized temperature $\theta$, degree of cure $\alpha$, and maximum temperature distributions for carbon/DCPD composite with $\phi=0.15$ obtained for (a) $T_0=$ 15$^oC$ and (b) $T_0=$ 30$^oC$. The increased effective thermal conductivity of the composite reduces the sharpness of the front and the amplitude of the thermal instabilities.}
	\label{composites_mix}
\end{figure}

For the fiber volume fraction $\phi=0.15$, we solve the set of PDEs described by (\ref{PDES_heat_sink}) to study the effects of convective heat loss on thermal instabilities associated with the FP-based manufacturing of carbon/DCPD composites. Figure \ref{composites_mix} presents snapshots of the temperature (blue curves) and degree-of-cure (red curves) distributions and spatial distribution of the maximum temperature (black curves) for $hP/S=110~kW/m^3.K$ and for $T_0=15^oC$ (Fig. \ref{composites_mix}(a)) and $T_0=30^oC$ (Fig. \ref{composites_mix}(b)). As apparent there, the increased thermal conductivity associated with the embedded carbon fibers leads to a less sharp thermal front and to reduced, less chaotic instabilities. At a higher temperature ($T_0=30^oC$), all instabilities are eliminated and a uniform spatial distribution of maximum temperature is obtained. 

\section{Conclusion}
Through a numerical analysis based on a thermo-chemical reaction-diffusion model, we have investigated the role of heat diffusion on the appearence of instabilities during the frontal polymerization of an adiabatic channel of neat DCPD resin and of unidirectional carbon/DCPD composites. The numerical study conducted in the neat resin case has shown how these FP-driven instabilities decay and the wavelength and amplitude of the thermal spike present in the front decrease as the initial temperature of the monomer increases, leading to an increasingly stable front propagation. While these thermal instabilities may influence the quality of the manufactured polymeric component, they do not affect the velocity of the polymerization front. 

In the case of carbon/DCPD composites, numerical results obtained with a homogenized thermo-chemical model have shown that instabilities may exist for lower values of the fiber volume fraction and initial temperature. A stable propagation of the polymerization front is achieved for a composite system at an initial temperature lower than that obtained in the neat resin case due to higher effective conductivity of the composite associated with presence of the embedded carbon fibers. The wavelength of the thermal instabilities increases with the fiber volume fraction, while the amplitude correspondingly decreases. 

Finally, we investigated the effect of convective heat losses on the front stability in neat resins and composites by adding a convective heat loss term to the thermo-chemical model. The presence of the convective heat loss led to two observations. Firstly, it extended the range of the initial temperature of the resin for which instabilities were obtained. Secondly, in contrast to the harmonic instabilities present in the adiabatic case, introducing a convective heat loss yielded substantially more complex, spatially varying instabilities, sometimes with chaotic patterns.

\section*{Acknowledgments}
This work was supported by the Air Force Office of Scientific Research through Award FA9550-16-1-0017 (Dr. B. “Les” Lee, Program Manager) as part of the Center for Excellence in Self-Healing, Regeneration, and Structural Remodeling. Authors also acknowledge the support of National Science Foundation for Grant No. 1830635, through the LEAP HI: Manufacturing USA Program.   
\bibliographystyle{ieeetr}
\bibliography{achs_jpcbfk122_4583}
\end{document}